\documentclass[aps,twocolumn,pra,superscriptaddress,showpacs,tightenlines]{revtex4}
\usepackage{epsfig,graphicx,times}
\usepackage{amstext}
\usepackage{amsmath}
\usepackage{amssymb}
\usepackage{graphicx}
\usepackage{latexsym}
\usepackage{bm}
\usepackage{CJK}

\begin{document}

\begin{CJK}{GBK}{song}
\title{Quantum anti-Zeno effect without wave function reduction}

\author{Qing Ai}
\affiliation{Advanced Science Institute, RIKEN, Wako-shi, Saitama 351-0198,
Japan} \affiliation{Institute of Theoretical Physics, Chinese
Academy of Sciences, Beijing, 100190, China}

\author{Dazhi Xu}
\affiliation{Advanced Science Institute, RIKEN, Wako-shi, Saitama
351-0198, Japan} \affiliation{Institute of Theoretical Physics,
Chinese Academy of Sciences, Beijing, 100190, China}

\author{Su Yi}
\affiliation{Advanced Science Institute, RIKEN, Wako-shi, Saitama
351-0198, Japan} \affiliation{Institute of Theoretical Physics,
Chinese Academy of Sciences, Beijing, 100190, China}

\author{A. G. Kofman}
\affiliation{Advanced Science Institute, RIKEN, Wako-shi, Saitama
351-0198, Japan} \affiliation{Physics Department, The University of
Michigan, Ann Arbor, Michigan 48109-1040, USA}

\author{C. P. Sun}
\affiliation{Advanced Science Institute, RIKEN, Wako-shi, Saitama
351-0198, Japan} \affiliation{Institute of Theoretical Physics,
Chinese Academy of Sciences, Beijing, 100190, China}

\author{Franco Nori}
\affiliation{Advanced Science Institute, RIKEN, Wako-shi, Saitama
351-0198, Japan} \affiliation{Physics Department, The University of
Michigan, Ann Arbor, Michigan 48109-1040, USA}

\date{\today}

\begin{abstract}

We study the measurement-induced enhancement of the spontaneous
decay (called quantum anti-Zeno effect) for a two-level
subsystem, where measurements are treated as couplings between the
excited state and an auxiliary state rather than the von Neumann's
wave function reduction. The photon radiated in a fast decay of the
atom, from the auxiliary state to the excited state, triggers a
quasi-measurement, as opposed to a projection measurement. Our use of
the term ``quasi-measurement" refers to a ``coupling-based measurement".
Such frequent quasi-measurements result in an exponential decay of the survival
probability of atomic initial state with a photon emission following each
quasi-measurement. Our calculations show that the effective decay
rate is of the same form as the one based on projection
measurements. What is more important, the survival probability of the
atomic initial state which is obtained by tracing over all the
photon states is equivalent to the survival probability of the
atomic initial state with a photon emission following each
quasi-measurement to the order under consideration. That is
because the contributions from those states with photon number less
than the number of quasi-measurements originate from higher-order
processes.



\end{abstract}
\maketitle \pagenumbering{arabic}
\end{CJK}

\section{Introduction}

\label{app:Intro}

In the quantum Zeno effect (QZE) (see, e.g., \cite%
{Namiki97,Koshinoa05,Degasperis74,Teuscher04,Khalhin68,Misra77}) frequent
measurements inhibit atomic transitions for a closed system. In the quantum
anti-Zeno effect (QAZE), atomic decays can be accelerated by frequent
measurements, when the observed atom also interacts with a heat bath with
some spectral distribution \cite%
{Lane83,Kofman00,Facchi01,Zheng08,Ai10,Ai10-2,Wang08,Zhou09,Cao10}. This
QAZE has been extensively studied for various cases, such as the QAZE
without the rotating-wave approximation \cite{Zheng08,Ai10,Cao10,Kofman04}
and in an artificial bath \cite{Ai10-2}. The conventional explorations for
the QAZE as well as the QZE need to invoke the von Neumann's wave function
collapse \cite{Neumann55} for quantum measurements, namely the projection
measurement postulate. Thus, the QAZE seems to depend on a particular
quantum mechanical interpretation specified by this collapse postulate.

However, even though the collapse postulate has been extensively used in the
past, some researchers do not believe it is necessary for quantum mechanics.
There exist other interpretations, such as the ensemble interpretation \cite%
{Ballentine70}. In this sense, it is necessary to develop a
quantum-mechanical-interpretation-independent approach to the QAZE.

To this end, we draw lessons from the dynamic explanations for the QZE \cite%
{Petrosky,Ballentine91,Block91,Frerichs91}. After the QZE was proposed by
Misra and Sudarshan \cite{Misra77}, it was recognized \cite{Peres80} that
the QZE could be mimicked by strong couplings to an external agent, which
carried out a coupling-based detection. Then, an experiment \cite{Itano90}
observing the QZE was explained \cite{Ballentine91} in such a dynamic
fashion. Therein, all the phenomena were only described by the unitary
evolution governed by the Schr\"{o}dinger equation for the whole system.
Later on, to further develop this dynamic interpretation of the QZE,
Pascazio \textit{et al.} \cite{Pascazio94} and Sun \textit{et al}. \cite%
{Sun94,Sun95} explicitly used the decoherence model of quantum measurement,
where the couplings to the apparatus only decohered the phases of the system
rather than changed the system's energy. This measurement model is
essentially a non-demolition measurement \cite{Braginsky92,Ashhab09}.

Following these dynamic approaches for the QZE, we now develop a quantum
dynamic theory for the QAZE without reference to projection measurements or
the collapse postulate. To illustrate our main idea, we use an example, a
two-level subsystem coupled to an auxiliary state to form a cascade
configuration. Due to the couplings to the reservoir, the excited state
spontaneously decays to the ground state. After a short interval, the
remaining population of the excited state is coherently pumped into the
auxiliary state by a strong laser. Then, it returns to the excited state by
a fast spontaneous decay and a photon is emitted simultaneously. At this
stage, a quasi-measurement is realized. Here, the term \textit{quasi-measurement}
refers to a \textit{coupling-based measurement} in contrast to the usual
projection measurement. The correlation of the atomic initial state and the
orthogonal states with two orthogonal states of the environment is produced
in such a process. We call it quasi-measurement since it can be viewed as the
first (unitary) stage of the measurement process. Similar to the conventional approach, based
on the collapse postulate, the effective decay rate of the survival
probability with one photon emitted following each pulse in the presence
of such quasi-measurements is given by the overlap
integral of the measurement-induced level-broadening function and the
interacting spectral distribution. As different photon states may not be
distinguished in a realistic experiment, the survival probability of the atomic
initial state after $n$ repetitive quasi-measurements, which can be obtained
by tracing over all the photon states, can be taken into consideration.
Since the contributions from photon states other than $n$ are due to
higher-order processes, they lead to a small correction to the final result
and can be omitted under the weak-coupling approximation in the short-time
regime. Thus, the result for the projection
measurements is recovered with the quasi-measurements.

The paper is organized as follows. In the next section, we describe the
physical setup to realize the \textit{dynamic} QAZE. In Sec.~\ref{app:AZEDV}%
, the effective decay rate of the survival probability with a photon emission
following each pulse is obtained for
repetitive quasi-measurements with a strong-intensity laser. The same result
is also attained for the survival probability of the atomic initial state. Finally,
we summarize the main results of the paper in Sec.~\ref{app:Con}. In order
to make the paper self-consistent for reading, we present the detailed
calculation for the free evolution under the short-time approximation in
Appendix~\ref{app:appendix1}.


\section{ Model Setup}

\label{app:MS}

\begin{figure}[tbp]
\includegraphics[scale=0.63]{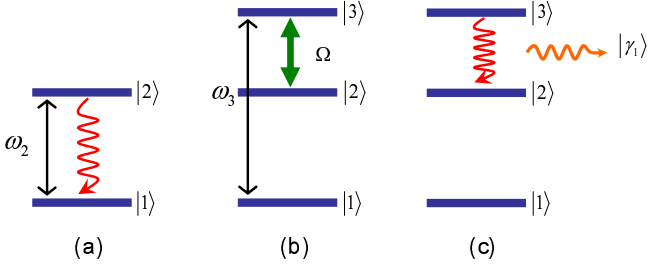}
\caption{(Color online) Energy level diagram for the three processes
considered here: (a) the spontaneous decay from the excited state $%
\left\vert 2 \right\rangle$ to the ground state $\left\vert 1 \right\rangle$%
, (b) a coherent transition with Rabi frequency $\Omega$ between $\left\vert
2 \right\rangle$ and the auxiliary state $\left\vert 3 \right\rangle$ by
laser pumping, and (c) a fast spontaneous decay from $\left\vert 3
\right\rangle$ to $\left\vert 2 \right\rangle$ with a photon emitted in $%
\left\vert \gamma_1 \right\rangle$. Here, the eigenenergies for the
excited state and the auxiliary state are $\omega_2$ and $%
\omega_3$, respectively.}
\label{sketch}
\end{figure}

We consider the QAZE for a three-level atom with the cascade configuration
depicted in Fig.~\ref{sketch}(b,c). We mainly focus on the QAZE concerning a
subsystem with the ground state $\left\vert 1\right\rangle $ and the excited
state $\left\vert 2\right\rangle $. Since these two levels are coupled to a
reservoir, there would be natural spontaneous decay from $\left\vert
2\right\rangle $ to $\left\vert 1\right\rangle $ if the subsystem were not
coupled to other dynamic agents. In this process with duration $\tau $, the
total system is governed by the Hamiltonian%
\begin{equation}
H=\sum\limits_{k}\omega _{k}a_{k}^{\dag }a_{k}+\omega _{2}\left\vert
2\right\rangle \left\langle 2\right\vert +\sum\limits_{k}g_{k}(a_{k}^{\dag
}\left\vert 1\right\rangle \left\langle 2\right\vert +\mathrm{H.c.})\text{,}
\label{H1}
\end{equation}%
where $a_{k}$ ($a_{k}^{\dag }$) is the annihilation (creation) operator for
the reservoir's $k$th mode with frequency $\omega _{k}$, $\omega _{2}$ the
eigenenergy for the excited state $\left\vert 2\right\rangle $, and $g_{k}$
the coupling constant between the $k$th mode and the transition between $%
\left\vert 1\right\rangle $ and $\left\vert 2\right\rangle $, which is
assumed to be real for simplicity. We assume $\omega _{1}=0$. Notice that we
have applied the rotating-wave approximation \cite{Scully97} to the above
Hamiltonian~(\ref{H1}).

\begin{figure}[tbp]
\includegraphics[scale=0.5]{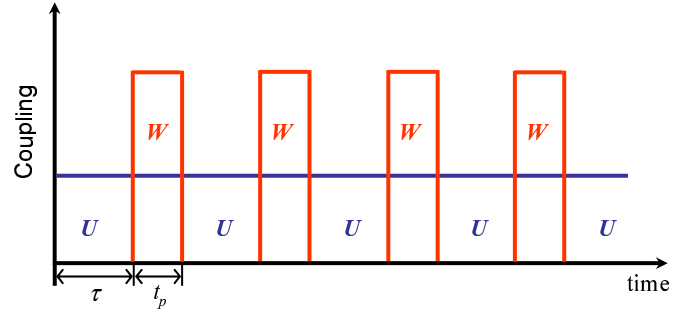}
\caption{(Color online) The pulse sequence for demonstrating the QAZE by a
quasi-measurement (i.e., avoiding projection measurements). Here, $U $ stands
for the spontaneous decay from $\left\vert 2\right\rangle $ to $\left\vert
1\right\rangle $. Also, $W$ is a quasi-measurement which is alternatively
present and absent for a duration $t_p$ and $\tau$,
respectively. }
\label{pulse}
\end{figure}

In order to perform a quasi-measurement, we avoid the collapse-postulate, as
also done, e.g., in Refs.~\cite{Peres80,Petrosky,Ballentine91}, where the
quasi-measurement involved coherently coupling the measured state to an
external agent, e.g., an additional energy level $\left\vert 3\right\rangle $.
In this sense, a quasi-measurement is the first (unitary) stage of the measurement
process, providing an entanglement between the system and the apparatus.
A quantum measurement in this approach is implemented by an alternative
coupling lasting for $t_{p}$ between $\left\vert 2\right\rangle $ and $%
\left\vert 3\right\rangle $ with eigenenergy $\omega _{3}$, which is
described by%
\begin{eqnarray}
H^{\prime } &=&\sum\limits_{k}\omega _{k}a_{k}^{\dag }a_{k}+\omega
_{2}\left\vert 2\right\rangle \left\langle 2\right\vert +\omega
_{3}\left\vert 3\right\rangle \left\langle 3\right\vert  \notag \\
&&+\Omega \cos \Delta t(\left\vert 2\right\rangle \left\langle 3\right\vert
+\left\vert 3\right\rangle \left\langle 2\right\vert )\text{,}  \label{Hp}
\end{eqnarray}%
where $\Omega $ is the Rabi frequency between $\left\vert 2\right\rangle $
and $\left\vert 3\right\rangle $. Hereafter, we focus on the resonance case,
i.e.,
\begin{equation}
\omega =\Delta \equiv \omega _{3}-\omega _{2}\text{.}
\end{equation}%
When the resonant coupling laser is applied between $\left\vert
2\right\rangle $ and $\left\vert 3\right\rangle $, we can disregard the
spontaneous decay between the auxiliary state $\left\vert 3\right\rangle $
and the excited state $\left\vert 2\right\rangle $\ for a very strong laser,
i.e., $\Omega \gg \Gamma $ with $\Gamma $\ being the decay rate from $%
\left\vert 3\right\rangle $\ to $\left\vert 2\right\rangle $. Then, when the
coupling laser is turned off, the population of the state $\left\vert
3\right\rangle $\ will quickly return to $\left\vert 2\right\rangle $, with
a photon $\gamma $ produced by the spontaneous decay, i.e.,
\begin{equation}
\left\vert 3,v\right\rangle \rightarrow \left\vert 2,\gamma
_{1}\right\rangle \text{,}  \label{3to2}
\end{equation}%
where $\left\vert 3,v\right\rangle =\left\vert 3\right\rangle \left\vert
v\right\rangle $ is the product state of the atomic auxiliary state $%
\left\vert 3\right\rangle $ and the vacuum $\left\vert v\right\rangle $ for
the reservoir, $\left\vert \gamma _{n}\right\rangle $ denotes the state with
$n$ photons in the $\gamma $ mode. At this stage, the quasi-measurement is
completed. Then, the subsystem alternatively evolves freely and is ``measured"
through laser pumping. The time sequence for the entire course is
schematically shown in Fig.~\ref{pulse}.

Here, we assume the duration $1/\Gamma $ for the fast spontaneous decay from
the auxiliary state $\left\vert 3\right\rangle $ to the excited state $%
\left\vert 2\right\rangle $ to be much smaller than the one for the
spontaneous decay from $\left\vert 2\right\rangle $ to $\left\vert
1\right\rangle $, i.e., $1/\Gamma \ll \tau$. In this case, we can omit the
dynamic evolution between $\left\vert 2\right\rangle $ and $\left\vert
1\right\rangle $ induced by the finite couplings to the reservoir when the
fast spontaneous decay from the auxiliary state $\left\vert 3\right\rangle $
to the excited state $\left\vert 2\right\rangle $ occurs.


\section{ Dynamical approach to the Anti-Zeno effect}

\label{app:AZEDV}

In the previous section, we described a dynamical approach to study the
QAZE. We emphasize that in our approach there is no wave-function-reduction
postulate involved, and the unitary evolution of both the two-level
subsystem and the measuring apparatus is depicted by means of the Schr\"{o}%
dinger equation. Let us first describe the two basic processes $U$ and $W$
schematically illustrated in Fig.~\ref{pulse}.

\subsection{ Spontaneous decay or $U$-process from $t=0$ to $\tau $}

For the spontaneous decay between the excited state and the ground state,
governed by the Hamiltonian (\ref{H1}), we assume the wave function of the
total system to be a superposition of two kinds of single-excitation states,
i.e.,%
\begin{equation}
\left\vert \Psi (t)\right\rangle =\alpha (t)\left\vert 2,v\right\rangle
+\sum\limits_{k}\beta _{k}(t)\left\vert 1,k\right\rangle \text{,}
\label{Psi}
\end{equation}%
where $\left\vert 2,v\right\rangle =\left\vert 2\right\rangle \left\vert
v\right\rangle $ is the product state of the atomic excited state $%
\left\vert 2\right\rangle $ and the vacuum $\left\vert v\right\rangle $ for
the reservoir, $\left\vert 1,k\right\rangle =\left\vert 1\right\rangle
\left\vert k\right\rangle $ the product state of the atomic ground state $%
\left\vert 1\right\rangle $\ and the single-excitation state $\left\vert
k\right\rangle $\ in the $k$th-mode of the reservoir. It follows from the
Schr\"{o}dinger equation $i\partial _{t}\left\vert \Psi (t)\right\rangle
=H\left\vert \Psi (t)\right\rangle $ that the coefficients $\alpha (t)$ and $%
\beta _{k}(t)$\ in Eq.~(\ref{Psi}) satisfy
\begin{subequations}
\label{EqAB}
\begin{eqnarray}
i\dot{\alpha} &=&\omega _{2}\,\alpha +\sum\limits_{k}g_{k}\,\beta _{k}\text{,%
} \\
i\dot{\beta}_{k} &=&\omega _{k}\,\beta _{k}+g_{k}\,\alpha \text{.}
\end{eqnarray}

Under the short-time approximation, the solutions to the above equations
become \cite{Kofman96}
\end{subequations}
\begin{subequations}
\label{AlphaBeta}
\begin{eqnarray}
\alpha (t) &=&\alpha (0)\left[ 1+\sum\limits_{k}g_{k}^{2}\frac{h_{t}(\omega
_{2}-\omega _{k})-it}{\omega _{2}-\omega _{k}}\right] e^{-i\omega _{2}t}
\notag \\
&&-\sum\limits_{k}g_{k}\beta _{k}(0)h_{t}(\omega _{2}-\omega
_{k})e^{-i\omega _{2}t}\text{,}  \label{Alpha} \\
\beta _{k}(t) &=&\beta _{k}(0)e^{-i\omega _{k}t}+\alpha (0)g_{k}h_{t}(\omega
_{2}-\omega _{k})e^{-i\omega _{2}t}\text{,}  \label{Beta}
\end{eqnarray}%
where
\end{subequations}
\begin{equation}
h_{t}(\omega )=\frac{1}{\omega }\left( e^{i\omega t}-1\right) \text{.}
\label{h}
\end{equation}%
The detailed calculations are presented in Appendix~\ref{app:appendix1}.

\subsection{ Quasi-measurement or $W$-process from $t=\tau $ to $%
\tau +t_{p}$}

In the quasi-measurement process, a strong laser field is applied to induce
the transition between the excited state $\left\vert 2\right\rangle $\ and
the auxiliary state $\left\vert 3\right\rangle $ [see Fig.~\ref{sketch}(b)].
With a unitary transformation%
\begin{equation}
W=\exp (-i\Delta t\left\vert 2\right\rangle \left\langle 2\right\vert )\text{%
,}
\end{equation}%
the transformed wave function $\left\vert \Psi ^{\prime }(t)\right\rangle
\equiv W\left\vert \Psi (t)\right\rangle $\ is governed by the effective
Hamiltonian $H_{\mathrm{eff}}\equiv WH^{\prime }W^{\dag }-iW\partial
_{t}W^{\dag }$, which reads%
\begin{equation}
H_{\mathrm{eff}}\simeq \sum\limits_{k}\omega _{k}a_{k}^{\dag }a_{k}+\omega
_{3}\left\vert 2\right\rangle \left\langle 2\right\vert +\omega
_{3}\left\vert 3\right\rangle \left\langle 3\right\vert +\frac{\Omega }{2}%
(\left\vert 2\right\rangle \left\langle 3\right\vert +\mathrm{H.c.})\text{,}
\end{equation}%
where we have dropped the fast-oscillating terms including the factors $\exp
(\pm i2\Delta t)$.

Now we assume the transformed wave function to be
\begin{equation}
\left\vert \Psi ^{\prime }(t)\right\rangle =A\left\vert 2,v\right\rangle
+\sum\limits_{k}B_{k}\left\vert 1,k\right\rangle +C\left\vert
3,v\right\rangle \text{.}
\end{equation}%
Then the original wave function $\left\vert \Psi (t)\right\rangle
=W^{-1}\left\vert \Psi ^{\prime }(t)\right\rangle $ can be written as%
\begin{equation}
\left\vert \Psi (t)\right\rangle =Ae^{i\Delta t}\left\vert 2,v\right\rangle
+\sum\limits_{k}B_{k}\left\vert 1,k\right\rangle +C\left\vert
3,v\right\rangle \text{.}
\end{equation}%
According to the Schr\"{o}dinger equation for the transformed wave function $%
i\partial _{t}\left\vert \Psi ^{\prime }(t)\right\rangle =H_{\mathrm{eff}%
}\left\vert \Psi ^{\prime }(t)\right\rangle $, we obtain the following
system of differential equations
\begin{subequations}
\label{EqABC}
\begin{eqnarray}
i\dot{A} &=&\omega _{2}A+\frac{\Omega }{2}Ce^{-i\Delta t}\text{,}
\label{EqAp} \\
i\dot{B}_{k} &=&\omega _{k}B_{k}\text{,}  \label{EqBkp} \\
i\dot{C} &=&\omega _{3}C+\frac{\Omega }{2}Ae^{i\Delta t}\text{.}
\label{EqCp}
\end{eqnarray}%
The solutions are given by
\end{subequations}
\begin{subequations}
\label{ABC}
\begin{eqnarray}
A(t) &=&\left[ A(0)\cos \frac{\Omega }{2}t-iC(0)\sin \frac{\Omega }{2}t%
\right] e^{-i\omega _{2}t}\text{,}  \label{Ap} \\
B_{k}(t) &=&B_{k}(0)e^{-i\omega _{k}t}\text{,}  \label{Bkp} \\
C(t) &=&\left[ C(0)\cos \frac{\Omega }{2}t-iA(0)\sin \frac{\Omega }{2}t%
\right] e^{-i\omega _{3}t}\text{.}  \label{Cp}
\end{eqnarray}

Applying a $\pi $-pulse, i.e., a laser with duration
\end{subequations}
\begin{equation}
t_{p}=\frac{\pi }{\Omega }\text{,}
\end{equation}%
drives the system to evolve into the state
\begin{equation}
\left\vert \Psi (t_{p})\right\rangle =\sum\limits_{k}B_{k}(t_{p})\left\vert
1,k\right\rangle +C(t_{p})\left\vert 3,v\right\rangle \text{,}
\end{equation}%
where the coefficients
\begin{subequations}
\begin{eqnarray}
B_{k}(t_{p}) &=&B_{k}(0)e^{-i\omega _{k}t_{p}}\text{,} \\
C(t_{p}) &=&-iA(0)e^{-i\omega _{3}t_{p}}
\end{eqnarray}%
can be obtained from Eq. (\ref{ABC}). Here, we have assumed there is no
initial population in the auxiliary state, namely $C(0)=0$, and thus $%
A(t_{p})=0$. Afterwards, by means of a fast spontaneous decay, the state $%
\left\vert 3,v\right\rangle $ decays into $\left\vert 2,\gamma
_{1}\right\rangle $ [see Fig.~\ref{sketch}(c)]. Therefore, a quasi-measurement
is finished.

\subsection{Repetition of the decay and quasi-measurement processes}

Here, we will explicitly describe the complete process including the free
evolution by $U$ and the quasi-measurement by $W$. The total system is
initially prepared in the excited state with the reservoir in the vacuum: $%
\left\vert \Psi (0)\right\rangle =\left\vert 2,v\right\rangle $. Then, after
a free evolution with period $\tau $, the state evolves into
\end{subequations}
\begin{equation}
\left\vert \Psi (\tau )\right\rangle =\alpha ^{(1,0)}(\tau )\left\vert
2,v\right\rangle +\sum\limits_{k}\beta _{k}^{(1,0)}(\tau )\left\vert
1,k\right\rangle \text{,}
\end{equation}%
where
\begin{subequations}
\begin{eqnarray}
\alpha ^{(1,0)}(\tau )\!\!\!\! &=&\!\!\!\!\left[ 1+\sum\limits_{k}g_{k}^{2}%
\frac{h_{\tau }(\omega _{2}-\omega _{k})-i\tau }{\omega _{2}-\omega _{k}}%
\right] e^{-i\omega _{2}\tau }\text{,} \\
\beta _{k}^{(1,0)}(\tau )\!\!\!\! &=&\!\!\!\!g_{k}h_{\tau }(\omega
_{2}-\omega _{k})e^{-i\omega _{2}\tau }\text{.}
\end{eqnarray}%
Applying a strong laser forces the system to evolve into
\end{subequations}
\begin{equation}
\left\vert \Psi (\tau +t_{p})\right\rangle =C(t_{p})\left\vert
3,v\right\rangle +\sum\limits_{k}\beta _{k}^{(1,0)}(\tau )\left\vert
1,k\right\rangle
\end{equation}%
with%
\begin{equation}
C(t_{p})=-i\alpha ^{(1,0)}(\tau )e^{-i\omega _{3}t_{p}}\text{.}
\end{equation}%
Later, through a fast spontaneous decay, the total system becomes
\begin{equation*}
\left\vert \Psi (\tau +t_{p}+t_{d})\right\rangle =C(t_{p})\left\vert
2,\gamma _{1}\right\rangle +\sum\limits_{k}\beta _{k}^{(1,0)}(\tau
)\left\vert 1,k\right\rangle \text{.}
\end{equation*}%
At this stage, the first cycle is accomplished. The survival probability
amplitude of the state $\left\vert 2\right\rangle $ after one
quasi-measurement is $C(t_{p})$. Hereafter, for the sake of simplicity, we
will label $\left\vert \Psi \left( n\tau +(n-1)t_{p}+(n-1)t_{d}\right)
\right\rangle $, $\left\vert \Psi \left( n\tau +nt_{p}+(n-1)t_{d}\right)
\right\rangle $, and $\left\vert \Psi \left( n\left( \tau
+t_{p}+t_{d}\right) \right) \right\rangle $ as $\left\vert \Psi
_{n}(1)\right\rangle $, $\left\vert \Psi _{n}(2)\right\rangle $, and $%
\left\vert \Psi _{n}(3)\right\rangle $, respectively. In other words, $%
\left\vert \Psi _{n}(j)\right\rangle $ denotes the state after $j$th
procedure in the $n$th cycle for $n=1,2,\cdots $ and $j=1,2,3$.

For the second cycle, after the free evolution, the total system is in the
state
\begin{eqnarray}
\left\vert \Psi _{2}(1)\right\rangle  &=&C(t_{p})\left[ \alpha ^{(1,0)}(\tau
)\left\vert 2\right\rangle +\sum\limits_{k}\beta _{k}^{(1,0)}(\tau
)\left\vert 1,k\right\rangle \right] \left\vert \gamma _{1}\right\rangle
\notag \\
&&+\alpha ^{(2,1)}(\tau )\left\vert 2,v\right\rangle +\sum\limits_{k}\beta
_{k}^{(2,1)}(\tau )\left\vert 1,k\right\rangle \text{,}
\end{eqnarray}%
where the coefficients
\begin{subequations}
\begin{eqnarray}
\alpha ^{(2,1)}(\tau ) &=&\sum\limits_{k}\left[ \beta _{k}^{(1,0)}(\tau )%
\right] ^{2}\text{,} \\
\beta _{k}^{(2,1)}(\tau ) &=&\beta _{k}^{(1,0)}(\tau )e^{-i\omega _{k}\tau }
\label{BetaK21}
\end{eqnarray}%
are determined by Eq.~(\ref{AlphaBeta}) with initial conditions
\end{subequations}
\begin{equation*}
\alpha ^{(2,1)}(0)=0\text{,}\;\;\;\beta _{k}^{(2,1)}(0)=\beta
_{k}^{(1,0)}(\tau )\text{.}
\end{equation*}%
After another $\pi $-pulse,%
\begin{eqnarray}
\left\vert \Psi _{2}(2)\right\rangle  &=&C(t_{p})\left[ \sum\limits_{k}\beta
_{k}^{(1,0)}(\tau )\left\vert 1,k\right\rangle +C(t_{p})\left\vert
3\right\rangle \right] \left\vert \gamma _{1}\right\rangle   \notag \\
&&+\sum\limits_{k}\beta _{k}^{(2,1)}(\tau )\left\vert 1,k\right\rangle
+C^{(2,1)}(t_{p})\left\vert 3,v\right\rangle \text{,}
\end{eqnarray}%
where
\begin{equation}
C^{(2,1)}(t_{p})=-i\sum\limits_{k}\left[ \beta _{k}^{(1,0)}(\tau )\right]
^{2}e^{-i\omega _{3}t_{p}}\text{.}  \label{C21}
\end{equation}%
Afterwards, by means of a fast spontaneous decay it becomes
\begin{eqnarray}
\left\vert \Psi _{2}(3)\right\rangle \!\!\!\! &=& \!\!\!\! C(t_{p})\sum\limits_{k}\beta
_{k}^{(1,0)}(\tau )\left\vert 1,k\right\rangle \left\vert \gamma
_{1}\right\rangle +\sum\limits_{k}\beta _{k}^{(2,1)}(\tau )\left\vert
1,k\right\rangle   \notag \\
\!\!\!\!&&\!\!\!\!+C^{2}(t_{p})\left\vert 2,\gamma _{2}\right\rangle
+C^{(2,1)}(t_{p})\left\vert 2,\gamma _{1}\right\rangle \text{.}
\end{eqnarray}

Thus, the survival probability of the state $\left\vert
2\right\rangle $ with photon emissions following both pulses is $%
\left\vert C^{2}(t_{p}) \right\vert^2$. Here, we point out that this is different from the
survival probability of the atomic initial state, which has an
additional contribution from $\left\vert 2,\gamma _{1}\right\rangle$.
In this dynamic approach for the QAZE, once a photon in the $\gamma $
mode is emitted right after a pulse, a quasi-measurement is finished. This means that the
system is in the initial state before the quasi-measurement and still remains
in its initial state after the quasi-measurement. For the case with two
quasi-measurements, $C^{(2,1)}(t_{p})$ corresponds to such a probability amplitude which
decays to the ground state before the first quasi-measurement and returns to the
excited state before the second quasi-measurement.

\subsection{ Survival probability describing the anti-Zeno effect}

By means of mathematical induction, we can prove the wave function of the
total system after $n$ quasi-measurements to be of the following form
\begin{eqnarray}
\left\vert \Psi _{n}(3)\right\rangle  &=&C^{n}(t_{p})\left\vert 2,\gamma
_{n}\right\rangle +\sum\limits_{j=1}^{n-1}C^{(n,j)}(t_{p})\left\vert
2,\gamma _{j}\right\rangle   \notag \\
&&+C^{n-1}(t_{p})\sum\limits_{k}\beta _{k}^{(1,0)}(\tau )\left\vert 1,k\right\rangle
\left\vert \gamma _{n-1}\right\rangle   \notag \\
&&+\sum\limits_{j=1}^{n-1}\sum\limits_{k}\beta _{k}^{(n,j)}(\tau )\left\vert
1,k\right\rangle \left\vert \gamma _{j-1}\right\rangle \text{,}
\end{eqnarray}%
where
\begin{equation}
C^{(n,j)}(t_{p})\sim \sum\limits_{k}\left[ \beta _{k}^{(1,0)}(\tau )\right]
^{2}  \label{Cnj}
\end{equation}%
for $j<n$. Judging from the analysis made in the previous section, we may
safely arrive at the conclusion that the survival probability amplitude of
the state $\left\vert 2\right\rangle $ with photon emissions following $n$
pulses is $C^{n}(t_{p})$. It is straightforward to calculate the survival
probability as%
\begin{equation}
P_{2}(t=n\tau )=\left\vert C(t_{p})\right\vert ^{2n}\text{.}  \label{P}
\end{equation}

As a result, we observe an exponential decay of the survival probability of
the atomic initial state with photon emission following each pulse,
i.e., $P_{2}(t)=\exp (-Rt)$. Here, the effective decay rate is%
\begin{equation}
R(\tau )=2\pi \int_{-\infty }^{\infty }F(\omega ,\tau )G(\omega )d\omega
\text{,}  \label{R}
\end{equation}%
where the interacting spectral distribution is%
\begin{equation}
G(\omega )=\sum\limits_{k}g_{k}^{2}\delta (\omega -\omega
_{k})=g_{k}^{2}\rho (\omega _{k})|_{\omega _{k}=\omega }\text{,}  \label{G}
\end{equation}%
with $\rho (\omega )$ being the density of state for $\omega $, the
measurement-induced level-broadening function
\begin{equation}
F(\omega ,\tau )=\frac{\tau }{2\pi }\text{sinc}^{2}\left[ \frac{1}{2}(\omega
_{2}-\omega )\tau \right] \text{.}
\end{equation}

\begin{figure}[tbp]
\includegraphics[scale=0.68]{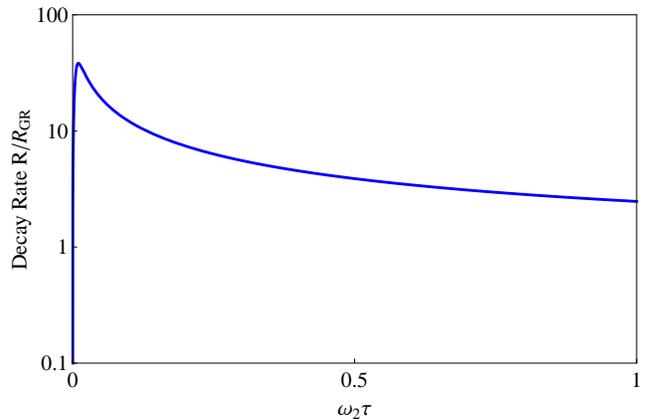}
\caption{(Color online) The effective decay rate $R(\tau)$ versus
quasi-measurement interval $\tau$ for the $2p$-$1s$ transition of the
hydrogen atom with $\omega_2=1.55\times 10^{16}$ rad/s.}
\label{RH2p1s}
\end{figure}

Besides, we resort to the numerical simulation for the $2p$-$1s$ transition
of hydrogen atom with the interacting spectral distribution \cite{Moses73}
\begin{equation}
G(\omega )=\frac{\eta \omega }{[1+(\frac{\omega }{\omega _{c}})^{2}]^{4}}%
\text{,}
\end{equation}%
where $\eta = 6.435\times 10^{-9}$, $\omega _{c}=8.491\times 10^{18}$ rad/s.
As shown in Fig.~\ref{RH2p1s}, the transition from the QAZE to the QZE,
which is the same as the one predicted by the projection measurement \cite%
{Kofman00}, is observed by varying the quasi-measurement interval $\tau$.
Here, the short interval for the QZE is roughly of the order of $1/\omega_c$
\cite{Facchi98}.

On the other hand, in a realistic experiment, on condition that the state $%
\left\vert 2,\gamma _{n}\right\rangle $ can not be distinguished from those
states $\left\vert 2,\gamma _{j}\right\rangle $ with $j$ less than $n$, the
survival probability of the initial state $\left\vert 2\right\rangle $ is
obtained by tracing over all the possible photon states. In this case, the
survival probability of the atomic initial state reads
\begin{equation}
P_{2}^{T}(t)=\left\vert C^{n}(t_{p})\right\vert
^{2}+\sum\limits_{j=1}^{n-1}\left\vert C^{(n,j)}(t_{p})\right\vert ^{2}\text{%
.}  \label{P2total}
\end{equation}

As seen in Eq.~(\ref{Cnj}), the contribution from the second term on the
right hand side of Eq.~(\ref{P2total}) is of higher-order correction to the
final result in the weak-coupling case. Thus, the survival probability of the atomic
initial state including the contributions from all the photon states also
displays an exponential decay with the effective decay rate being of the
same form in Eq.~(\ref{R}). This is a physical result and its reason will be
presented as follows. After the $n$th cycle, $C^{(n,j)}(t_{p})$ is the
probability amplitude for $\left\vert 2,\gamma _{j}\right\rangle $, which
stands for $j$ photons emitted in $n$ quasi-measurements. Take $%
C^{(n,n-1)}(t_{p})$ for an example. It corresponds to such a probability
amplitude which decays into the ground state before one quasi-measurement and
returns to the initial state before the next quasi-measurement. For the other $%
n-2$ quasi-measurements, it always stays in the initial state before the
quasi-measurements. Since the transition probability amplitude between $%
\left\vert 2\right\rangle $ and $\left\vert 1,k\right\rangle $ is $g_{k}$,
the probability amplitude for the atomic excited state to first decay into
the ground state and then return is proportional to $g_{k}^{2}$. As a
result, the corresponding probability $\left\vert
C^{(n,n-1)}(t_{p})\right\vert ^{2}$ is of the order of $g_{k}^{4}$. Similar
analyses can also be applied to $\left\vert C^{(n,j)}(t_{p})\right\vert ^{2}$
when $j<n-1$. In a word, the second term on the right hand side of Eq.~(\ref%
{P2total}) can be neglected due to its characteristics of higher-order
correction.

Since the above analysis is based on the weak-coupling approximation, a
natural question comes into our minds. That is under what condition the
second term on the right hand side of Eq.~(\ref{P2total}) can be
disregarded. A straightforward calculation shows
\begin{equation}
\left\vert \sum\limits_{k}\left[ \beta _{k}^{(1,0)}(\tau )\right]
^{2}\right\vert ^{2}\ll \text{Re}\sum\limits_{k}\frac{g_{k}^{2}h_{\tau
}(\omega _{2}-\omega _{k})}{\omega _{2}-\omega _{k}} \label{Cond}
\end{equation}%
with Re$(x)$ being the real part of $x$. Here, the term on the right hand side of Eq.~(\ref{Cond})
corresponds to the second order term in Eq.~(\ref{P}), while the term on the left hand side of Eq.~(\ref{Cond})
corresponds to the second term on the right hand side of Eq.~(\ref{P2total}). Mathematically speaking, the
interacting spectrum $G(\omega )$ should be sufficiently broad and smooth.
This is similar to the quantitative criterion obtained for the spontaneous
decay to a continuum with photonic band gaps \cite{Kofman94}. In the case
with strong couplings, the return of the excitation from the final state may
be significant for sufficiently-long times as already shown in Refs.~\cite%
{Milburn88,Kofman01}, where the free evolution was due to a classical field.

\section{Conclusion}

\label{app:Con}

In this paper, we investigated the QAZE for a two-level subsystem embedded
in a three-level atom. Instead of considering projection measurements, we
studied quasi-measurements by pumping the population of the excited state to
an auxiliary state. Since the pumped population returns to the excited state
by a fast spontaneous decay, the complete process of the quasi-measurement is
finished. Along with the fast spontaneous decay, there is a photon emitted
in the corresponding mode.

We found that the effective decay rate of the survival probability still
remains as the overlap integral of the measurement-induced level-broadening
function and the interacting spectral distribution. Moreover, it is
discovered that the survival probability of the atomic initial state is the same as the
survival probability of the atomic initial state with photon emission following each
pulse since the difference between them leads to a higher-order correction. This is because the
contributions from the other photon states originate from higher-order processes. In conclusion,
without projection measurements, we can observe the QAZE and the QZE by
means of quasi-measurements.

Generally speaking, the QZE and QAZE stem from frequent decoherence events,
which destroy the off-diagonal density matrix elements. When the diagonal elements
in the density matrix remain unchanged after these
processes, the above decoherence is actually dephasing between the initial
and final states, e.g., between the first and second terms on the right hand
side of Eq.~(\ref{Psi}). And the model in this paper is just of such kind.
Other methods include measurements as in Refs.~\cite%
{Pascazio94,Sun94,Sun95,Milburn88,Facchi05}, and even a classical random
field \cite{Harel98}. Note that the above decoherence can take effect due to
not only dephasing, but also a destruction of the final states \cite%
{Kwiat98,Kofman01}. On the other hand, the decoherence can be suppressed
by a train of ultrafast off-resonant optical pulses \cite{Search00}.


\begin{acknowledgments}
We thank Adam Miranowic and X. F. Cao for useful comments. FN
acknowledges partial support from the National Security Agency, Laboratory
of Physical Sciences, Army Research Office, National Science Foundation
grant No. 0726909, JSPS-RFBR contract No. 09-02-92114, Grant-in-Aid for
Scientific Research (S), MEXT Kakenhi on Quantum Cybernetics, and FIRST
(Funding Program for Innovative R\&D on S\&T). C. P. Sun is supported by the
NSFC Grant No. 10935010. QA is supported
by China Postdoctoral Science Foundation grant No. Y0Y2301B11-10B101.
\end{acknowledgments}


\appendix

\section{Time Evolution in Spontaneous Decay}

\label{app:appendix1}

In this Appendix, we present the detailed calculations for the free
evolution. By defining the slowly-varying variables%
\begin{eqnarray}
\alpha ^{\prime }(t) &=&\alpha e^{i\omega _{2}t}\text{,}  \label{transAlpha}
\\
\beta _{k}^{\prime }(t) &=&\beta _{k}e^{i\omega _{k}t}\text{,}
\label{transBeta}
\end{eqnarray}%
we obtain a system of simplified equations from Eq.~(\ref{EqAB})%
\begin{eqnarray}
i\dot{\alpha}^{\prime } &=&\sum\limits_{k}g_{k}\beta _{k}^{\prime
}e^{i(\omega _{2}-\omega _{k})t}\text{,}  \label{EqAlpha} \\
i\dot{\beta}_{k}^{\prime } &=&g_{k}\alpha ^{\prime }e^{-i(\omega _{2}-\omega
_{k})t}\text{.}  \label{EqBeta}
\end{eqnarray}%
We can integrate Eq.~(\ref{EqBeta}) to have a formal solution for $\beta
_{k}^{\prime }(t)$, i.e.,
\begin{eqnarray}
\beta _{k}^{\prime }(t) &=&\beta _{k}^{\prime
}(0)-i\int\limits_{0}^{t}g_{k}\alpha ^{\prime }(t^{\prime })e^{i(\omega
_{k}-\omega _{2})t^{\prime }}dt^{\prime }  \notag \\
&\simeq &\beta _{k}^{\prime }(0)-i\alpha ^{\prime
}(0)\int\limits_{0}^{t}g_{k}e^{i(\omega _{k}-\omega _{2})t^{\prime
}}dt^{\prime }  \notag \\
&=&\beta _{k}(0)-\alpha (0)g_{k}h_{t}(\omega _{k}-\omega _{2})\text{,}
\label{BetaP}
\end{eqnarray}%
where in the second line we have used the short-time approximation $\alpha
^{\prime }(t^{\prime })\simeq \alpha ^{\prime }(0)$ and $h_{t}(\omega )$
given by Eq. (\ref{h}). By substituting Eq.~(\ref{BetaP}) into Eq.~(\ref%
{EqAlpha}) and making use of the short-time approximation, we have
\begin{eqnarray}
\alpha ^{\prime }(t)\!\!\! &=&\!\!\!\alpha ^{\prime
}(0)-i\sum\limits_{k}\int\limits_{0}^{t}g_{k}\beta _{k}^{\prime }(t^{\prime
})e^{-i(\omega _{k}-\omega _{2})t^{\prime }}dt^{\prime }  \notag \\
\!\!\! &\simeq &\!\!\!\alpha (0)\left[ 1+\sum\limits_{k}g_{k}^{2}\frac{%
h_{t}(\omega _{2}-\omega _{k})-it}{\omega _{2}-\omega _{k}}\right]  \notag \\
\!\!\! &&\!\!\!-\sum\limits_{k}g_{k}\beta _{k}(0)h_{t}(\omega _{2}-\omega
_{k})\text{.}  \label{AlphaP}
\end{eqnarray}

In order to obtain the explicit forms of $\alpha $\ and $\beta _{k}$, we
shall use the inverse transform of Eqs.~(\ref{transAlpha}) and (\ref%
{transBeta}),%
\begin{eqnarray}
\alpha (t) &=&\alpha (0)\left[ 1+\sum\limits_{k}g_{k}^{2}\frac{h_{t}(\omega
_{2}-\omega _{k})-it}{\omega _{2}-\omega _{k}}\right] e^{-i\omega _{2}t}
\notag \\
&&-\sum\limits_{k}g_{k}\beta _{k}(0)h_{t}(\omega _{2}-\omega
_{k})e^{-i\omega _{2}t}\text{,}
\end{eqnarray}%
and $\beta _{k}(t)$ given by Eq. (\ref{Beta}).


\end{document}